\documentclass[10pt]{article}
\usepackage{amsmath}
  \usepackage{paralist}
  \usepackage{graphics} 
  \usepackage{epsfig} 
 \usepackage[colorlinks=true]{hyperref}
\hypersetup{urlcolor=blue, citecolor=red}

\usepackage{amsmath, amsthm, amssymb}

\title {Boolean Models of Bistable Biological Systems}

\author{Franziska Hinkelmann and Reinhard Laubenbacher}

\begin{document}
\maketitle


\medskip

\centerline{\scshape Franziska Hinkelmann$^{a,b,c}$ and Reinhard
Laubenbacher$^{b,c}$}
\medskip
{\footnotesize 
 \centerline{$^a$Interdisciplinary Center for Applied Mathematics,
    Virginia Polytechnic Institute and State University,} 
  \centerline{Blacksburg, VA 24061-0531, USA}
}
{\footnotesize
 \centerline{$^b$Department of Mathematics,
   Virginia Polytechnic Institute and State University,}
   \centerline{Blacksburg, VA 24061-0123, USA}
}
{\footnotesize
 \centerline{$^c$Virginia Bioinformatics Institute,
    Virginia Polytechnic Institute and State University,}
 \centerline{Blacksburg, VA 24061-0477, USA}
}

\bigskip


\begin{abstract}
This paper presents an algorithm for approximating certain types of dynamical
systems given by a system of ordinary delay differential equations by
a Boolean network model. Often Boolean models are much simpler to understand than
complex differential equations models. The motivation for this work comes
from mathematical systems biology. While Boolean mechanisms do not provide information about exact
concentration rates or time scales, they are often sufficient to
capture steady states and other key dynamics. Due to their intuitive nature, such
models are very appealing to researchers in the life sciences. 
This paper is focused on dynamical systems that exhibit bistability and are described
by delay equations. It is shown that if a certain motif
including a feedback loop is present in the wiring diagram of the system, the
Boolean model captures the bistability of molecular switches. The method is applied to
two examples from biology, the lac operon and the phage $\lambda$
lysis/lysogeny switch. 
\end{abstract}
\section{Introduction}
Since the discovery of the first gene regulatory network, the lactose
metabolism network in the bacterium \emph{E. Coli} by Jacob Monod,
\cite{Monod:1961}, such networks have been modeled extensively, traditionally
with differential equations. But other modeling techniques like Boolean
networks \cite{Albert:topology}, stochastic models \cite{Burrage:2006}, Petri
nets, or Bayesian networks \cite{Friedman:2000} have also been used
successfully. Work by Kauffman in 1969 \cite{Kauffman:1969} suggested that
gene networks behave like Boolean switching nets and therefore Boolean
networks are suitable to model them. Using a Boolean network model, it has
been shown, for instance, that in some cases the network topology and the
logic of the interactions among the different molecular species is sufficient
to determine the qualitative dynamics of the network, without taking into
account the detailed kinetics \cite{Albert:topology}. In cases where not
enough information about the kinetic parameters is available to build a
detailed continuous model, one often still has enough information to build a
Boolean network model which can provide important information. 

The purpose of this paper is to show for a particular family of continuous models
that they can be approximated by a Boolean network model that retains the
key information about model dynamics. Our intent is to demonstrate that
Boolean models can be used to study a variety of structural and dynamic 
features of biological and other systems that have traditionally been modeled
using continuous models. They have the added advantage that they are intuitive
and do not require much mathematical background. This makes them particularly
suitable for use in the life sciences. The focus of this paper is on two
features: bistability and time delays. We give a general algorithm how to
approximate a dynamical system given by a system of ordinary delay
differential equations by a Boolean network model and validate it with two
well-known examples from systems biology. The relation between discrete and
continuous models has been examined before, e.g., \cite{Edwards:symbolic}
describes how a logical network can be used to
create the analogous differential equations. Other methods allow to
systematically create a continuous model from a discrete dynamical system
\cite{Mendoza:2006}, or to reduce the model space by using a
finite state machine \cite{Conradi:2001}. We briefly
review the main concepts related to Boolean network models. 
\subsection{Introduction to Boolean Models}\label{introbool}
We use a time discrete deterministic Boolean network with synchronous update.
In a Boolean model, a variable can
only be in the ``on'' (1) or ``off'' (0) state. In systems biology applications, each variable
typically corresponds to a molecular entity, e.g., the concentration of a gene product
such as mRNA or proteins.
A Boolean network of $n$ variables
consists of an update function 
$f = (f_1, \ldots, f_n): \mathbb F _2 ^n \rightarrow \mathbb F _2 ^n $, also
called a transition function. 
The dynamics of the system are generated deterministically by iteration of the transition
function $f$. 
It can be visualized by its phase space, which
shows all possible states and their transitions. A cycle in the phase space is
called a limit cycle; if it has length one, a fixed point. Fixed points of a
discrete system are the equivalent to steady states of a continuous system. 

In the dependency graph, also referred to as wiring diagram, each variable is a
vertex and an edge from variable $x_i$ to $x_j$ is drawn if $x_i$ shows up in
the local transition function $f_j$. A directed cycle in the dependency graph is
called a feedback loop. 
\subsubsection*{Example}
\begin{figure}[htp]
	\begin{center}
	\includegraphics[width=.28\columnwidth]{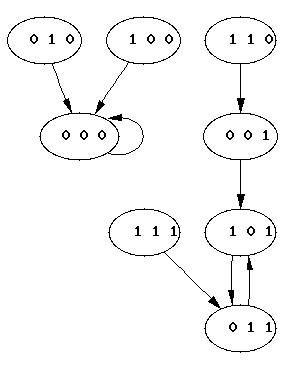}
	\caption{\textit{Phase space of $f = (\neg x_1 \wedge x_3, x_1 \wedge x_3,
	(x_1 \wedge x_2) \vee x_3) $ }}
	\label{fig:example}
	\end{center}
\end{figure}

Consider the following Boolean network with three variables: 
\begin{align*}
f_1 &= ((\neg x_1) \wedge x_3)\\
f_2 &= (x_1 \wedge x_3)\\
f_3 &= ((x_1 \wedge x_2) \vee x_3).
\end{align*}
The phase space of this function is depicted in Fig \ref{fig:example}. The
network has a limit cycle of length two because the state $(1\ 0\ 1)$
transitions into $(0\ 1\ 1)$, which turns back to $(1\ 0\ 1)$. It also
includes the fixed point $(0\ 0\ 0)$.
Discrete Visualizer of Dynamics (DVD tool) \cite{dvd} was used to create the phase
space. 
It calculates the
number and length of limit cycles and fixed points, as well as the
trajectories from any initial state. It also generates a graph of the phase
space along with the dependency graph.
\section{Approximation by Discrete Models}

We begin by describing several relevant features of biochemical networks
that a model needs to capture and describe in each case how this is done
in the Boolean network case. We use lower case letters for continuous
variables and upper case letters for discrete ones. 

\subsubsection{Dilution and Degradation}\label{sec:dilution}
It is common in biochemical networks that the concentration of a substance $X$
decreases with time because of dilution and degradation. In a differential
equations model this is usually accounted for with a negative degradation term: 
$$\frac{dx}{dt} = \cdots - \gamma x.$$
In the discrete model, for simplicity it is assumed that the degradation rate
for a substance is either vanishingly small and can therefore be ignored, or
the substance is degraded below the threshold for discretization after two
discrete timesteps. 

To model the decrease of concentration of a substance by dilution and
degradation, which is modeled in the continuous case by a negative degradation
term for $x$, a new variable $X_{\text{old}}$ is introduced. If
$X_{\text{old}}$ is ``on,'' then any quantity of $X$ that is available has
already been reduced once by dilution and degradation, and if no new substrate
is produced, $X$ will be ``off'' in the next time step. 
\subsubsection{Time Delay}\label{sec:tau}
In gene regulatory networks, time delays are often caused by transcription and
translation. 
Dependence of a variable $x$ on the concentration of a substance $y$ time $\tau$ ago
 can be described with a delay differential equations model
$$\frac{d}{dt}x(t)= \ldots \ y(t-\tau) \ \ldots$$
In a time discrete model, further
variables are needed to model this delay.
One has to choose the length $\overline{t}$ of a discrete timestep. If the
delay $\tau =
\overline{t}$, then $X$ is chosen to depend on $Y_1$ and $Y_1$ depends on $Y$. 
\begin{align*}
f_X &= \ldots Y_1 \ldots\\
f_Y &= \ldots \\
f_{Y1} &= Y
\end{align*}
If $\tau > \overline t$, additional variables have to be
introduced. 
\subsubsection{Distinguishing between Low, Medium, and High Concentrations}
To differentiate between three states for a variable $X$ in a Boolean system,
a second variable $X_{\text{high}}$ is introduced. By doing so one can make a
distinction between low, medium, and high concentration of $X$: if $X$ is on,
the concentration is a least medium, whereas $X_{\text{high}}$ is only on, if the
concentration is high.
\subsubsection{Bistability}
If previous experiments or computations have shown that the system
exhibits bistable behavior depending on the concentration of a stimulus
$X$, it is necessary to distinguish in the model between different
concentrations of $X$: the concentrations for which there is a unique
steady state for the system and for which there are multiple possible
steady states. Without this distinction, a Boolean model could not
capture bistability. In the lac operon, a medium allolactose
concentration leads to bistability, therefore the discrete model 
must differentiate between low, medium, and high concentration of
allolactose.
\begin{figure}[htp]
	\begin{center}
	\includegraphics[width=.23\columnwidth]{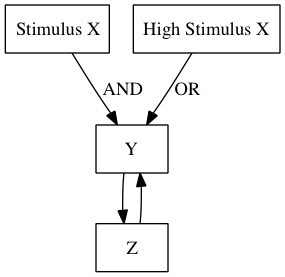}
	\caption{\textit{Necessary motif in dependency graph for bistability}}
	\label{fig:bistability}
	\end{center}
\end{figure}
The stimulus $X$ is the parameter that drives a system in and out of the
bistable region; within the bistable region the previous state determines the
resulting steady state. The stimulus acts on one or more variables. Usually
$X_{\text{high}}$ is part of an OR statement in the transition function (if the
stimulus is present in high concentration, it should overrule everything and
turn the system on), whereas $X$ is part of an AND statement (under its
presence other factors might prevent the system from turning on). The state of
$Z$ (this could be a set of variables or a single variable) indicates
whether the system was in a preinduced state before the stimulus concentration
was changed or not, depending on the input from 
$Z$: $$Y = (Z \wedge X) \vee X_{\text{high}}.$$
To assure that $Z$ reflects whether the system was preinduced, $Z$ needs to
receive input from $Y$. This leads to a system that can be reduced to a system
as the one shown in Fig.
\ref{fig:bistability}.
As long as the dependency graph of a Boolean model can be reduced to Fig.
\ref{fig:bistability}, it is guaranteed, that the discrete network captures
the bistability of the system correctly.
\subsection{Example of Building Boolean Model}
\begin{figure}[htp]
	\begin{center}
	\includegraphics[width=\columnwidth]{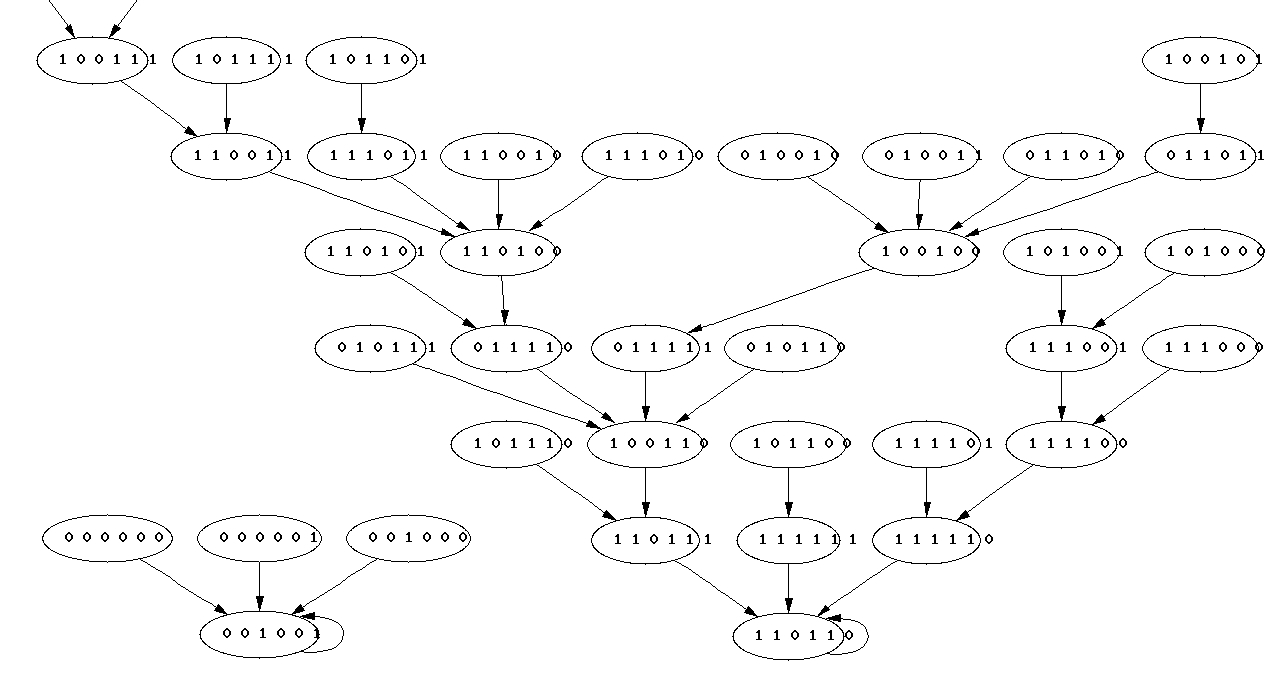}
	\caption{\textit{Phase space of network \ref{discr_equ}} }
	\label{fig:example_ps}
	\end{center}
\end{figure}

We demonstrate the above method on the following generic delay differential equations
model:
\begin{align*}
  \frac {dx} {dt} &= y(t - \tau) - 1 \tag{1} \label{cont_equ}\\
  \frac {dy} {dt} &= x(t - \tau) -  1.
\end{align*}
The steady state of this system is $(x,y) = (1,1)$. When translating
\ref{cont_equ} into a
discrete model, the first step is to generate Boolean equations for $x$ and
$y$, 
\begin{align*}
  f_x &= y\\
  f_y &= x.
\end{align*}
To account for the degradation term $-1$ in the equation for $y$, a new
variable
$y_{\text{old}}$ is introduced as outlined in section \ref{sec:dilution}, and,
similarly, $x_{\text{old}}$.
We use $x_1$ and $y_1$ for the delay $\tau$ ,
as described in section \ref{sec:tau}. This leads to the following system
\begin{align*} 
  f_x &= y_1 \vee ( x \wedge \neg x_{\text{old}})  \\
  f_{x_1} &= x\\
  f_{x_\text{old}} &= \neg y_1.\\
  f_y &= x_1 \vee ( y \wedge \neg y_{\text{old}}) \tag{2} \label{discr_equ}\\
  f_{y_1} &= y\\
  f_{y_\text{old}} &= \neg x_1.
\end{align*}
Part of the phase space of the network is depicted in Fig. \ref{fig:example_ps}. It clearly shows $(x,y) = (1,1) = (1,\_\_,1,\_,\_)$ as
its fixed point, just as expected from the solution of the continuous model.
In the
continuous system values for $x$ and $y$ could be negative, this is
represented in the fixed point $(x,y) = (0,0) = (0,\_\_,0,\_,\_)$ of
the discrete system.

In the next two sections we apply the above methods to two well-studied examples
of gene regulatory networks, for which there exist delay differential equations
models. 
\section{The \textit{Lac} Operon}
The lac operon is required for the transport and metabolism of lactose in
\textit{Escherichia coli}. It has been studied extensively, and
Novick \cite{Novick:1957} showed in the
1950s that bistability is observed with artificial inducers. He observed that ``preinduced'' bacteria, that is bacteria grown in a high concentration of inducer, are able to
maintain a high internal inducer concentration if subsequently grown
in a low external inducer concentration. Novick, \cite{Novick:1957}
calls this the ``preinduction effect''. He observed, however, that if
preinduced bacteria are transferred to a medium with no inducer,
enzyme synthesis ceases immediately. The minimum concentration in
which the high rate of synthesis of a preinduced culture can be
maintained is called the \textit{maintenance concentration}. 

Novick's experiments have led to various mathematical models of the lac operon whose
steady state solutions show bistability, for example Yildirim
\cite{Yildirim:feedback}, Wong\cite{Wong-lacoperon}, Boer \cite{boer:tb}.

\begin{figure}[htp]
	\begin{center}
	\includegraphics[width=.8\columnwidth]{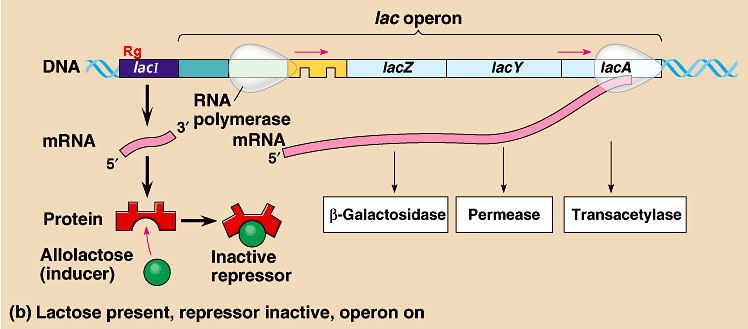}
	\caption{\textit{Schematic representation of the lactose operon
  regulatory system from dnainfo.wikispaces.com} }
	\label{fig:lacoperon}
	\end{center}
\end{figure}

We first give a brief overview of the functionality of the lac operon
depicted in Fig. \ref{fig:lacoperon}. In the
absence of glucose for cellular metabolism, extracellular lactose is
transported into the cell, either actively by permease or passively through
diffusion. Inside the cell, $\beta$-galactosidase breaks up lactose into
glucose, galactose, and allolactose. Allolactose binds to the repressor, which
is usually bound to the operator region where it inhibits the transcription
process, therefore the transcription process can proceed. RNA polymerase
initiates transcription of the structural genes to produce mRNA, which is then
translated into proteins, permease, and $\beta$-galactosidase.

The system contains a positive feedback loop, with an increase in
the concentration of allolactose, mRNA concentration increases, and therefore
more $\beta$-galactosidase and permease are produced, which in turn leads to
more internal lactose and allolactose.
Positive (or negative) feedback
loops are very common in biological systems because they allow for a
rapid increase in concentration. In the case of the lac operon, the enzyme
induction is an ``all-or-none phenomenon'', with the first
permease molecule extracellular lactose is transported into the
cell, increasing the inducer concentration, and therefore increasing
the probability of the appearance of another permease molecule,
\cite{Novick:1957}.
\subsection{Boolean Model of Lac Operon}
We derive a Boolean network for the gene regulatory network from the
continuous model developed in \cite{Yildirim:feedback} consisting of 5
equations. All its parameters were estimated from the biological literature
and the equations were numerically solved for their steady states. The delay
differential equations model predicts a bistable region for a medium external
inducer concentration, which is in accordance to what has been observed
experimentally by 
\cite{Novick:1957} and \cite{Cohn:1959}. As in
the continuous model, we take the following 5 variables for the Boolean model into
account: messenger RNA $M$, $\beta$-galactosidase $B$, allolactose $A$, permease
$P$, and internal lactose $L$. 

We will show how to derive the Boolean equation for messenger RNA, the process
for the other variables is similar and can be found in the appendix. In
\cite{Yildirim:feedback}, the rate of change of mRNA is given by the following
equation:
\begin{align*}
\frac {d}{dt} M(t)
 =& \alpha_M \frac {1+K_1(e^{-\mu \tau_M}A_{\tau_M})^n}
{K + K_1(e^{-\mu \tau_M}A_{\tau_M} )^n} + \Gamma_0 - \tilde\gamma_M
M\notag\\
 =& \alpha_M \frac {1+K_1(e^{-\mu \tau_M} A(t-\tau _ M))^n}
{K + K_1(e^{-\mu \tau_M} A(t-\tau _ M) )^n} + 
\Gamma_0 - (\gamma_M + \mu)M,
\end{align*}
where $n$ is the number of molecules of allolactose required to
inactivate the repressor.
$\frac {d}{dt}M(t)$ depends on $A$ at time $(t - \tau_M)$ and $-(\gamma_M + \mu)M$ models the
loss caused by dilution and degradation. 

Under the presence of allolactose $A$, $\frac {d}{dt}M$ is non-negative, so
the Boolean equation for mRNA is
\begin{align*}
f_M = x_{A_{\tau}},
\end{align*}
where $A_{\tau}$ describes the allolactose concentration $A$ time $\tau_M$
ago. 

If mRNA is present and no new mRNA is produced in the next time step, the
concentration will decrease according to the degradation rate. To capture this
in the Boolean model, we introduce the artificial variable $M_{\text{old}}$ as
described in section \ref{sec:dilution}. $M_{\text{old}}$ is ``on'', if
allolactose is absent, because then no new mRNA was produced. If no mRNA was
produced for 2 time steps, its concentration is too low and we consider $M$ to
be ``off''. This results in the following equations: 
\begin{align*}
f_M &= x_{A_{\tau}} \vee (x_M \wedge \neg x_{M_{\text{old}}})\\
f_{M_{\text{old}}} &= \neg x_{A_{\tau}}.
\end{align*}

mRNA depends on $A$ time $\tau$ ago, so a new variable $A1$ is introduced, as
described in \ref{sec:tau} and we set 
\begin{align*}
f_M &= x_{A1} \vee (x_M \wedge \neg x_{M_{\text{old}}})\\
f_{M_{\text{old}}} &= \neg x_{A1}\\
f_{A} &= \ldots \\
f_{A1} &= x_A.
\end{align*}

\begin{figure}[htp]
	\begin{center}
	\includegraphics[width=.5\columnwidth]{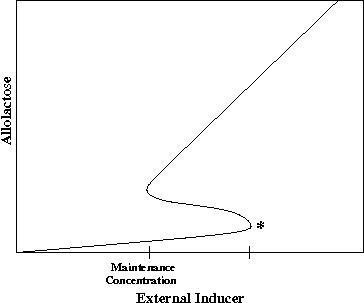}
	\caption{\textit{Discretization of External Inducer}}
	\label{fig:discretization}
	\end{center}
\end{figure}

We discretize the external inducer concentration $L_e$ to be ``on'', if it is
above the minimal maintenance concentration. The high external inducer
concentration $L_{e_{\text{high}}}$ is ``on'', if the concentration is at
least
(*), as shown in Fig \ref{fig:discretization}. 

Using the method described above, accounting for degradation and delays, we
derive the following Boolean model: \\
\begin{minipage}{.45\textwidth}
\begin{align*}
&f_M = x_{A1} \vee (x_M \wedge \neg x_{M_{\text{old}}})\\
&f_{M1} = x_{M}\\
&f_{M2} = x_{M1}\\
&f_{M3} = x_{M2}\\
&f_{M_{\text{old}}} = \neg x_{A1}\\
&f_B = x_{M2} \vee x_B\\
&f_A = (x_B \wedge x_L) \vee (x_L \wedge x_{L_{e_{\text{high}}}}) \\
&   \qquad \vee (x_A \wedge \neg x_{A_{\text{old}}})\\
&f_{A1} = x_A\\
\end{align*}
\end{minipage}
\begin{minipage}{.45\textwidth}
\begin{align*}
&f_{A_{\text{old}}} =\neg x_B \vee \neg x_L\\
&f_L = (x_P \wedge x_{L_e}) \vee x_{L_{e_{\text{high}}}} \\
&   \qquad \vee [x_L \wedge \neg( x_{L_{\text{old}}} \wedge (x_P \vee x_B))]\\
&f_{L_{\text{old}}} = \neg x_P \vee \neg x_{L_e}\\
&f_P = x_{M3} \vee (x_P \wedge \neg x_{P_{\text{old}}})\\
&f_{P_{\text{old}}} = \neg x_{M3}\\
&f_{L_e} = x_{L_e} \vee x_{L_{e_{\text{high}}}}\\
&f_{L_{e_{\text{high}}}} = x_{L_{e_{\text{high}}}}.\\
\end{align*}
\end{minipage}\\
We use the DVD simulation software, see section \ref{introbool}, to calculate that the system has
8 fixed points and no limit cycles.

As expected, a low external inducer concentration (last two variables are
set to 0) drives the system to fixed points corresponding to the operon in
the ``off'' state, points $1-3$ in table \ref{table:fp}. If the concentration
of the artificial external inducer is medium, but not high, the system results
in the fixed points $4-7$ in table \ref{table:fp}. With high external inducer
concentration, the system settles in the remaining fixed point which corresponds
to an induced operon. 
\begin{table}
\hspace{-.45in}
\begin{tabular}{c|ccccccccccccccc}
&$M$ &$M_1$ & $M_2$ & $M_3$ & $M_{\text{old}}$ & $B$ & $A$ & $A_1$ &
$A_{\text{old}}$ &$ L$ & $L_{\text{old}}$ & 
$P$ &$P_{\text{old}}$ & $L_e$ & $L_{e_{\text{high}}}$\\
\hline
1& 0&0&0&0&1&0&0&0&1&0&1&0&1&0&0\\
2& 0&0&0&0&1&0&0&0&1&1&1&0&1&0&0\\
3& 0&0&0&0&1&1&0&0&1&0&1&0&1&0&0\\
\hline
4& 0&0&0&0&1&0&0&0&1&1&1&0&1&1&0\\
5& 0&0&0&0&1&1&0&0&1&0&1&0&1&1&0\\
6& 0&0&0&0&1&0&0&0&1&0&1&0&1&1&0\\
7& 1&1&1&1&0&1&1&1&0&1&1&1&0&1&0\\
\hline
8& 1&1&1&1&0&1&1&1&0&1&1&1&0&1&1\\
\end{tabular}
\caption{Fixed Points of Lac Operon}
\label{table:fp}
\end{table}

For a medium inducer concentration, fixed point $4$ and $5$ are biologically
not meaningful, because they correspond to states in which only the internal
lactose or $\beta$-galactosidase concentration is present but no other
substance. Fixed point $6$ corresponds to the ``off'' operon, fixed point $7$
to the ``on'' operon.

To show bistability, we analyze the behavior of the fixed points as we change
the concentration rates of the external inducer. If we start
in a state corresponding to fixed point $1$ and increase the inducer
concentration to a medium concentration, the system settles down in fixed
point $6$, corresponding to the ``off'' operon. If we start with state 
$8$, the system settles down in fixed point $7$, corresponding to the ``on''
operon. This is exactly what we expect from the solution of the delay
differential equation: under a medium concentration of external inducer, the
steady state depends on whether the cell was preinduced or not.
\section{Lambda Phage}
\begin{figure}[htp]
	\begin{center}
	\includegraphics[width=.8\columnwidth]{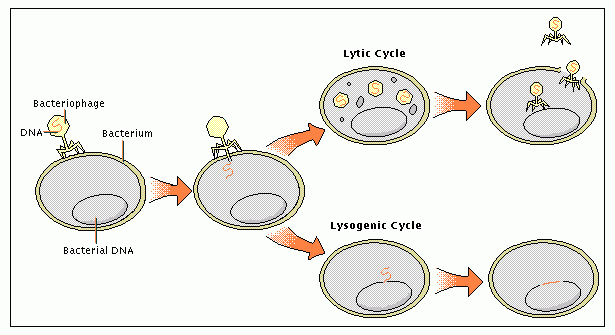}
	\caption{\textit{The Life Cycle of Lambda Phages from www.mining.ubc.ca} }
	\label{fig:phagelambda}
	\end{center}
\end{figure}

The virus lambda phage (phage $\lambda$) is a bacteriophage that infects
\textit{Escherichia coli}. After injecting its DNA into the host, the phage can enter
the lytic pathway where it alters the host DNA to produce phage particles and
then lyses the host cell, or it can enter the lysogenic pathway, where it is
duplicated with every cell division and harmless until the cell is under
stress, then it enters its lytic pathway. A schematic representation of
the phage $\lambda$ switch is shown in Fig. \ref{fig:phagelambda}.
Interestingly, the lysogenic state
is more stable than the genome itself \cite{Sneppen:2002}. A comprehensive explanation of
the lambda phage switch can be found in \cite{Ptashne:2004}.

In \cite{Santillan:2004} a delay differential equations model for the switch
between lysogenic and lytic state of phage lambda is presented. This continuous
model consists of four equations for the rate of change of the concentrations
of \textit{cI} and \textit{cro} messenger RNA molecules, $cI$ and Cro monomers
and of two equations for the dimer concentrations of $[CI_2]$ and $[Cro_2]$. 

Bistable behavior depends on the degradation rate of $cI$, $\gamma _{cI}$.
Numerical calculations show that the range for $\gamma _{cI}$ in which the
systems has two fixed points, one corresponding to lysogeny, the other to
lysis, is $0.0 \text{min}^{-1} < \gamma _{cI} < 0.35 \text{min}^{-1}$. If the
degradation rate of $cI$ is zero, the phage enters the lysogenic
pathway, if it is above $0.35 \text{min}^{-1}$, the lytic pathway. For a
medium degradation rate of $cI$, phage lambda can enter the lytic or lysogenic
pathway, so one needs to distinguish between three different degradation
rates of $\gamma _{\text{cI}}$. Note that for the lambda phage a degradation rate
drives the system in the bistable region, whereas for the lac operon an
external inducer has that role. To distinguish between three states of $\gamma
_{\text{cI}}$ in the Boolean model, the variable $\gamma _{\text{cI}_{high}}$ is
introduced. To avoid wrong input, e.g., $\gamma _{\text{cI}} = 0$ and $\gamma
_{\text{cI}_{high}}=1$, $\gamma _{\text{cI}}$ is turned on, whenever $\gamma
_{\text{cI}_{high}}=1$.

The delay caused by transcription is only 0.005 $\text{min}^{-1}$,
whereas the delays caused by the translation of the monomers $cI$ and Cro are
0.06 and 0.24 $\text{min}^{-1}$. Therefore delay caused by transcription is
neglected in the model. 

For a high degradation rate, no $cI$ monomers are left after 1 timestep. 
We assume, that a small concentration of monomers is generated also if the
Boolean expression that generates $cI$ is zero. As a consequence, if the 
degradation rate for $cI$ is zero, we assume that sufficient $cI$ monomers are
produced to set $x_{CI}=1$ after one time step. 

This results in the following Boolean model. 
\begin{align*}
f_{M_{\text{cI}}} &= x_{CI _ T} \vee \overline{x_{Cro _ T}} \vee
		(x_{M_{\text{cI}}} \wedge \overline
	{ x_{M_{\text{cI}_{\text{old}}}}})\\
f_{M_{\text{cI}_{\text{old}}}} &= \overline{x_{CI _ T}} \wedge x_{Cro _ T}\\
		x_{M_{\text{cI}_1}} &= x_{M_{\text{cI}}} \\
f_{M_{\text{cro}}} & = \overline{x_{CI _ T}} \vee (x_{M_{\text{cro}}} \wedge
	\overline{ x_{{M_{\text{cro}}} _{\text{old}}} }) \\
f_{{M_{\text{cro}}} _{\text{old}}} &= x_{CI _ T}\\
		x_{M_{\text{cro}_1}} &= x_{M_{\text{cro}}}\\
		x_{M_{\text{cro}_2}} &= x_{M_{\text{cro}_1}}\\
f_{CI _ T} &= \overline{\gamma_{\text{cI}_1}} \vee (\overline{\gamma_{\text{cI}_
		\text{high}}} \wedge (x_{M_{\text{cI}_1}} \vee (x_{CI _ T} \wedge
		(\overline{ x_{{CI _ T}_{\text{old}}}} \vee \overline{\gamma_{\text{cI}} }))))\\
f_{{CI _ T}_{\text{old}}} &= \overline{x_{M_{\text{cI}_1}} } \vee \gamma_{\text{cI}_ \text{high}}\\
f_{Cro _ T} & = x_{M_{\text{cro}_2}} \vee
	(x_{Cro _ T} \wedge \overline{ x_{{Cro _ T} _{\text{old}}}})\\
f_{{Cro _ T} _{\text{old}}} &= \overline{ x_{M_{\text{cro}_2}}}\\
f_ {\gamma_{\text{cI}}} &= \gamma_{\text{cI}} \vee \gamma_{\text{cI}_ \text{high}}\\
f_{\gamma_{\text{cI}_1}} &= \gamma_{\text{cI}} \\
f_{\gamma_{\text{cI}_ \text{high}}} &= \gamma_{\text{cI}_ \text{high}}.
\end{align*}

DVD simulation software \cite{dvd} is used to calculate the fixed points and generate the
dynamics in the bistable region. If the degradation rate is $0.0
\text{min}^{-1}$, in the model the last three variables are off, the system
results in the fixed point $[ 1 0 1 0 1 0 0 1 0 0 1 0 0 0 ]$ which corresponds
to $[ M_{cI} CI]$, a high $cI$ and low Cro concentration which means that the
system is in the lysogenic state. If the degradation rate is medium, in the
model the third last variable is on and the last two variables are off, there
are two fixed points, $[ 0 1 0 1 0 1 1 0 1 1 0 1 1 0 ]$ and $[ 1 0 1 0 1 0 0 1
0 0 1 1 1 0 ]$, corresponding to $[ M_{cro} Cro]$ and $[ M_{cI} CI]$,
respectively, lysogenic and lytic pathway. With a high degradation rate,
in the model only the last variable of the last three variables is on, the
system settles in the fixed point $[ 0 1 0 1 0 1 1 0 1 1 0 1 1 1 ] $,
representing $[ M_{cro} Cro]$. This means that the system is in the lytic
state. All four fixed points are in accordance with the results found numerically 
and the switching behavior observed experimentally.

To investigate the bistable region, we start the model with an initialization that
represents the lysogenic state. Increasing the degradation rate to a
medium level results in the fixed point $[1 0 1 0 1 0 0 1 0 0 1 1 1 0 ]$,
which is still the lysogenic state. Starting the model from a lytic state and
decreasing the rate results in the fixed point $[ 0 1 0 1 0 1 1 0 1 1 0 1 1 0]$, representing the lytic
state. Decreasing the rate even further finally results in the fixed point
corresponding to the lysogenic state. 
\section{Discussion}
The results presented in this paper show that biochemical networks that
exhibit bistability can be modeled successfully using a Boolean network
model, incorporating delays for variables, as needed. This was done by
showing that continuous delay-differential equations models can be
approximated by Boolean networks. The method presented here is quite
general and could be applied to other types of biological networks. 
The examples show that simple Boolean
models are able to capture steady states and complicated dynamics like hysteresis. 

Boolean network models have the drawback that they do not give rise to exact concentration
rates of the steady states because their discretization is too coarse grained.
Discrete models with more than just two states, so called multi state models,
might be suitable to give enough quantitative information about concentration
rates while they are intuitive enough and easy to use for a wide range of scientists. 

\section*{Acknowledgements}We would like to especially thank Terry Herdman for
his invaluable support and encouragement the first author without which this research could
not have been possible. 

\begin{appendix}
\section{\textit{Lac} Operon}
\subsection{Dilution and Degradation}
Since the degradation rates $\gamma_M$, $\gamma_P$, and $\gamma_A$ are close to
$0.5\text{min}^{-1}$, we 
assume that mRNA, permease, and allolactose are degraded after 2 times
steps. 
$\gamma_L$ and $\gamma_B$ are very small and will be neglected in our
model. 
\subsection{$\beta$-galactosidase}
For the $\beta$-galactosidase enzyme the equation in the continuous model is 
\begin{align*}
\frac{dB}{dt} &= \alpha_B e^{-\mu \tau_B}M_{\tau_B} - \tilde\gamma_B
B\notag\\
 &= \alpha_B e^{-\mu \tau_B}M_{\tau_B} - (\gamma_B + \mu) B.
\end{align*}
Messenger RNA is translated into $\beta$-galactosidase which takes time
$\tau_B$, so $f_B$
depends on 
$M_{\tau}$, the mRNA concentration time $\tau_B$ ago.

Since the degradation rate $\gamma_B$ is low, the Boolean model neglects the
decrease due to dilution and degradation and we model $\beta$-galactosidase
with the single equation
\begin{align*}
f_B &= x_{M_{\tau}} \vee x_B.
\end{align*}
\subsection{Allolactose $A$}
\begin{align*}
\frac{dA}{dt} &= \alpha_A B \frac L {K_L +L} - \beta_AB \frac A {K_A + A}
- \tilde\gamma_A A\\
\end{align*}
Allolactose is gained by conversion of lactose and reduced by the
loss via conversion to glucose and galactose, both mediated by
$\beta$-galactosidase. Like for mRNA, we use an extra variable $A_{\text{old}}$
to capture the loss of allolactose due to dilution and degradation.
\begin{align}\label{discreteA}
f_A &= (x_B \wedge x_L) \vee (x_A \wedge \neg x_{A_{\text{old}}})\\
f_{A_{\text{old}}} &= \neg (x_B \wedge x_L) \notag\\
&= \neg x_B \vee \neg x_L \notag
\end{align}
Notice that \ref{discreteA} does not depend on any delayed variables. 
\subsection{Internal Lactose $L$}
For internal lactose, Yildirim's model suggests the following equation
\begin{align*}
\frac {dL}{dt} = &\alpha_LP \frac {L_e}{K_{L_e} + L_e} - \beta_LP \frac L
{K_{L_1} + L} \notag\\
&- \beta_{L_2}B \frac L {K_{L_2} + L} - \tilde \gamma_L L.
\end{align*}
The degradation term $\gamma_L$ is low enough to be ignored in the discrete
model. Lactose is broken down into glucose, galactose and allolactose by
$\beta$-galactosidase, and permease actively transports lactose in and out of the
cell. Again, since permease and $\beta$-galactosidase reduce the internal
lactose concentration, we introduce the extra variable
$L_{\text{old}}$ to turn $L$ off, if it has not been produced and if permease
or $\beta$-galactosidase are present to reduce it. 
\begin{align*}
f_L &= (x_P \wedge x_{L_e}) \vee [x_L \wedge \neg( x_{L_{\text{old}}}
\wedge (x_P \vee x_B))]\\
f_{L_{\text{old}}} &= \neg(x_P \wedge x_{L_e})\\
&= \neg x_P \vee \neg x_{L_e}\\
\end{align*}
\subsection{Permease $P$}
\begin{align*}
\frac {dP}{dt} &= \alpha _P e ^{-\mu(\tau_P + \tau_B)}M_{\tau_P +
\tau_B} - \tilde \gamma _p P \label{Yildirim2}\\
\end{align*}
Messenger RNA is translated into permease, so the 
permease concentration $P$ is directly proportional to the mRNA concentration $M$ at time
$(\tau_P + \tau_B)$ ago and dilution and degradation reduce permease
concentration, which is why $P_{\text{old}}$ is used in the Boolean model.
\begin{align*}
f_P &= x_{M_{\tau}} \vee (x_P \wedge \neg x_{P_{\text{old}}})\\
f_{P_{\text{old}}} &= \neg x_{M_{\tau}}.
\end{align*}
\section{Lambda Phage}
\begin{figure}[htb]
  \includegraphics[width=.3\columnwidth]{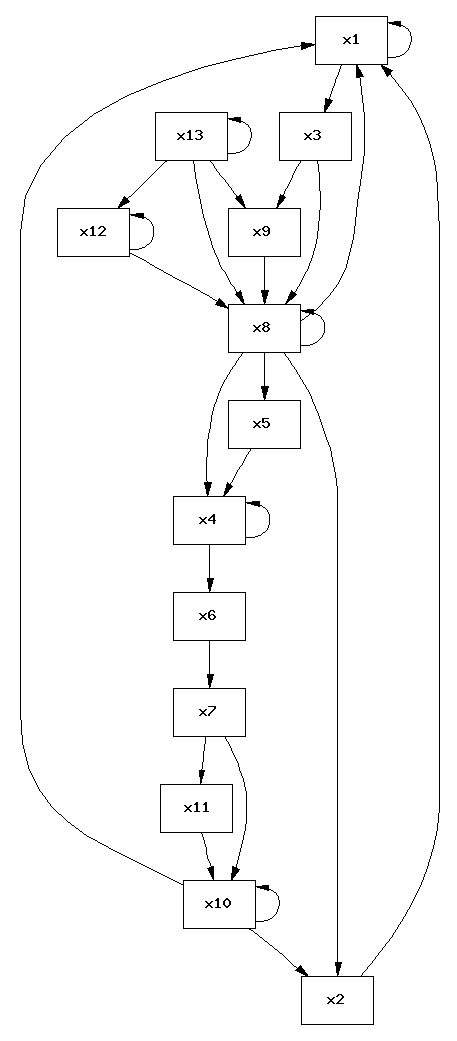}
  \caption{Dependency graph of the Boolean model for Lambda Phage}
  \label{fig:lambda}
\end{figure}

Renumerating the equations results in 
\begin{align*}
f_1 &= x_8 \vee \overline{ x_{10}} \vee (x_1 \wedge \overline { x_2}) \\
f_2 &= \overline{x_8} \wedge x_{10}\\
f_3 &= x_1\\
f_4 & = \overline{x_8} \vee (x_4 \wedge
	\overline{x_5})\\
f_5 &= x_8\\
f_6 &= x_{4}\\
f_7 &= x_6\\
f_8 &= \overline{x_{13}} \vee (\overline{x_{14}} \wedge (x_{3} \vee
	(x_8 \wedge (\overline{ x_9} \vee \overline{x_{12}}))))\\
f_9 &= \overline{x_3 } \vee x_{14}\\
f_{10} & = x_7 \vee
	(x_{10} \wedge \overline{ x_{11}})\\
f_{11} &= \overline{ x_7}\\
f_{12} &= x_{12} \vee x_{14}\\
f_{13} &= x_{12}\\
f_{14} &= x_{14}.
\end{align*}
The Boolean model has the dependency graph shown in Fig. \ref{fig:lambda}. 
\end{appendix}
\bibliographystyle{amsplain}
\bibliography{discretemodel}

\end{document}